\documentclass[11pt]{article}
\usepackage[utf8]{inputenc}

\usepackage{amsmath}
\usepackage{amssymb}
\usepackage{amsthm}
\usepackage{authblk}

\usepackage{algpseudocode}
\usepackage{algorithm}
\usepackage{tikz}
\usetikzlibrary{positioning}

\usepackage{fullpage}
\usepackage{xcolor}
\usepackage{hyperref}
\usepackage[multiple]{footmisc}

\title{A Brief Note on a Recent Claim About NP-Hard Problems and~BQP\thanks{This brief note was prepared for the Theorietag~2024 Workshop on Algorithms, Complexity, and Logic at Heinrich-Heine-Universit\"{a}t D\"{u}sseldorf,  following a request from J\"{o}rg Rothe, who we thank for this opportunity.}~${}^,$\thanks{Work supported in part by NSF grant CCF-2006496.}}

\author{Michael C. Chavrimootoo} 
\affil{Department of Computer Science\\University of Rochester\\Rochester, NY 14627, USA}

\newcommand{\naturalnumber}{\ensuremath{{\mathbb{N}}}}

\newcommand{\p}{\ensuremath{{\rm P}}}
\newcommand{\np}{\ensuremath{{\rm NP}}}

\newtheorem{innercustomthm}{Theorem}
\newenvironment{customthm}[1]
  {\renewcommand\theinnercustomthm{#1}\innercustomthm}
  {\endinnercustomthm}

\date{April 25, 2024}

\begin{document}\sloppy

\maketitle

\begin{abstract}
This short note outlines some of the issues in Czerwinski's paper~\cite{cze:t:np-hard-bqp} claiming that NP-hard problems are not in BQP\@.
We outline one major issue and two minor issues, and conclude that their paper does not establish what they claim it does.
\end{abstract}

\section{Overview}

In their paper, Czerwinski attempts to show that BQP contains no NP-hard set~\cite{cze:t:np-hard-bqp}. 
One natural consequence of such a result would be a separation of $\p$ and $\np$.
However, there are several flaws in the argument. In the next section, we outline one major flaw and two minor ones (among others) in that paper. The major flaw we find is enough to invalidate their main ``proof,'' i.e., that of their Theorem~2 (reproduced later in this note). Indeed, their purported proof uses a false assumption; they attempt to prove that a particular process is not computable and from that conclude their result, but that aforementioned process is not just computable, it is computable in exponential time! Moreover, while their Theorem~2's statement is unclear, the natural way to interpret that statement implies that their Theorem~2 is false.

\section{Preliminaries}

Let $\naturalnumber = \{0,1,2,\ldots\}$ denote the set of natural numbers.
In this note, we assume standard familiarity with undecidability, recursive many-one reductions, oracle Turing machines (OTMs), and NP, as  described in standard textbooks such as~\cite{hop-ull:b:automata,aro-bar:b:complexity,sip:b:introduction-third-edition}.

\section{Main Analysis}

Beyond several issues related to poor notation or to lack of clarity, there are three key pitfalls of the paper that are highlighted below. 
Because we reproduce two ``theorems'' of Czerwinski in this note, we use the same theorem numbering used by their paper in order to eliminate any confusion when referring to ``Theorem~1'' or  to ``Theorem~2.''
We first give the minor issues as they are relevant to further critiquing the major issue.

\subsection{First Minor Issue}
In their Section~3~\cite{cze:t:np-hard-bqp}, Czerwinski defines for an arbitrary Turing machine (TM) $M$ the set
$$D_M = \{1^n \mid (\exists x \in \{0,1\}^n)[x \in L(M)]\}.$$

They then claim that the ``set $D_M$ is not computable for an arbitrary TM $M$.''
That is a fundamental misunderstanding in mathematical terminology. The statement is equivalent to saying that ``for every TM $M$, the set $D_M$ is not computable,'' which is clearly false whenever $L(M)$ is decidable.\footnote{An analogy to their flawed argument is as follows: ``Not every natural number is composite as some natural numbers are prime. Thus an arbitrary natural number is prime.''}

It seems that they want to argue ``it cannot hold that for each TM $M$, $D_M$ is computable,'' which is true, but not equivalent to their statement quoted in the previous paragraph. This issue becomes relevant in the next two issues highlighted below.

\subsection{Second Minor Issue}
Using that flawed argument from the previous subsection, they attempt to prove the following result, which is false.

Consider their Theorem~1, which is reproduced below with some minor edits for clarity.

\begin{customthm}{1}[\cite{cze:t:np-hard-bqp}]
Let $M$ be an arbitrary TM\@. A quantum computer or OTM with oracle $L(M)$ cannot decide whether $1^n \in D_M$ faster than with a black box search.
\end{customthm}

The notion of ``black box search'' is not clearly defined in Czerwinski's paper, but the above claim seems to imply that any oracle machine with oracle $L(M)$ that decides $D_M$ must make at least one oracle call. For an arbitrary machine $M$, this is clearly false as $L(M)$ could be decidable. And if $L(M)$ is undecidable, that is trivial since it's easy to see that $L(M) \leq_m D_M$, and an OTM with oracle $L(M)$ that makes no call to the oracle and yet decides $D_M$ would imply that $L(M)$ is decidable (a contradiction).
Thus, even under the restriction that $L(M)$ is undecidable in Theorem~1, this gives us no new insight.

This ``theorem'' is used later in the proof of Theorem~2.

\subsection{The Major Issue}

The approach of Czerwinski resembles one used in an earlier paper by the same author (see~\cite{cha-le-rei-smi:t:czerwinski} for a summary and critique of that earlier paper).

For a given natural number $t$ and machine $M$, Czerwinski seems to use $M_t$ to denote a machine that on input $x$ simulates $M$ on input $x$ for $t$ steps.
For a machine $M$, the set $U_M$ is defined as
$$\{(1^n, 1^t) \mid (\exists x \in \{0,1\}^n)[x \in L(M_t)]\}.$$

\begin{customthm}{2}[\cite{cze:t:np-hard-bqp}]
    Let $n,t \in \naturalnumber$ be arbitrary but fixed and $M$ an arbitrary TM\@. Testing whether $(1^n, 1^t) \in U_M$ cannot be done faster than with a black box search.
\end{customthm}

The exact meaning of Theorem~2's statement is not clear either, but the most natural interpretation seems to be that they are claiming that no deterministic machine that decides $U_M$ can be ``oracle-free.''
In fact, their purported proof that BQP contains no NP-hard problems seems to rely on this interpretation.
But, because they put no time bound on a machine accepting $U_M$ and since $U_M \in \np$ (this is given as their Lemma~2, but it is also easy to see), it follows that there is a deterministic exponential-time Turing machine that accepts $U_M$, thus disproving their Theorem~2.

Giving them the benefit of doubt, we can further look into their purported proof (this is somewhat a ``sub-issue'' of the major issue). Their purported proof states:

\begin{quote}
[if] one could test whether $(1^n, 1^t) \in U_M$ faster than with black box search [...] [then] one could decide if there is such an $x$. But that is undecidable.
\end{quote}

That statement is clearly false; to compute a string of length $n$ that is accepted by $M$ within $t$ steps, simply iterate over each (binary) length-$n$ string $y$ (there are finitely many such strings) and output $y$ if $M$ accepts $y$ within $t$ steps. Clearly, this is an exponential-time-computable process---there are $2^n$ binary strings of length $n$ and for each binary string $x$ of length $n$, checking if $M$ accepts $x$ within $t$ steps can be done in time polynomial in $n+t$---thus contradicting Czerwinski's claim.

So, we have that the purported  proof of Theorem~2 uses a false assumption, so Czerwinski's paper fails to prove Theorem~2.
Additionally, under the above interpretation, which seems to be the most natural one, Theorem~2 is false.

The subsequent claims in the paper (that $\np \not\subseteq {\rm BQP}$, that BQP contains no NP-hard set, and that $\p \neq \np$) supposedly follow from Theorem~2, which Czerwinski's paper fails to prove. Thus that paper also fails to establish those three subsequent claims.

\section{Conclusion}

We did not dive into the other results stated in~\cite{cze:t:np-hard-bqp} as those seem to be known results that are often assigned as exercises in courses on computability and complexity theory. 
Nonetheless, because we did not verify them, we do not claim that they are correct. We can however conclude that Czerwinski's paper~\cite{cze:t:np-hard-bqp} fails to prove what it claims.

\bibliographystyle{alpha}
\bibliography{gry-reu,local_refs}

\end{document}